# The Search for Relevance: A Context-Aware Paradigm Shift in Semantic and Task-Oriented V2X Communications

Luca Lusvarghi, *Member, IEEE*, Javier Gozalvez, *Fellow, IEEE*, Baldomero Coll-Perales, *Member, IEEE,* Mohammad Irfan Khan, *Member, IEEE,* Miguel Sepulcre, *Senior Member, IEEE,* Seyhan Ucar, *Senior Member, IEEE* and Onur Altintas, *Member, IEEE*

***Abstract*—The design of communication systems has traditionally prioritized the reliable and timely delivery of data. However, the scalability challenges faced by the evolution towards a data-driven hyper-connected society and economy demand new communication paradigms that carefully curate the content being transmitted. This paper proposes a joint semantic and task-oriented communication paradigm where connected devices transmit only the information necessary to convey the desired meaning that is relevant to the intended receivers, based on their context. We qualitatively and quantitatively analyze the potential benefits of the proposed semantic and task-oriented communication paradigm in the Vehicle-to-Everything (V2X) domain. The V2X domain offers a unique environment for the development and deployment of semantic and task-oriented V2X communications, as it is rich in contextual information and Connected and Autonomous Vehicles (CAVs) are native semantic devices. The qualitative analysis focuses on a cooperative perception use case and shows how semantic and task-oriented V2X communications can reduce the amount of information transmitted by each vehicle without compromising the situational awareness of its intended receivers. The quantitative analysis numerically demonstrates that semantic and task-oriented V2X communications can achieve a two-fold improvement in communication efficiency which can significantly benefit the scalability of future V2X networks.**

## I. INTRODUCTION

THE advent of Vehicle-to-Everything (V2X) communications will drive the digital transformation of transportation systems and pave the way for a safer, more efficient, and more sustainable Cooperative, Connected, and Automated Mobility (CCAM). With V2X communications, Connected and Autonomous Vehicles (CAVs) will be able to exchange information through direct or network-based communications [1], enhancing their situational awareness, perception, and cooperative driving capabilities [2].

Like any existing communication system, current V2X communications prioritize the reliable and timely transmission of messages, addressing the technical communication problem identified by Shannon and Weaver at Level A [3]: *"how accurately can the communication symbols be transmitted?"*. Level A is the first of three levels (A, B, and C) defined by Shannon and Weaver to systematically organize and analyze the broad subject of communication. Level A communication systems are agnostic to the content of the transmitted messages and can result in the transmission of unnecessary information, potentially compromising communication efficiency and network scalability. Efficiency and scalability will be critical challenges in 6G and beyond, as the connectivity demand and the number of connected devices continues to grow. These challenges are particularly critical in the V2X domain, where V2X-enabled CCAM applications will require CAVs to continuously broadcast rich information. Current V2X networks are not ready to handle the large increase in data traffic that will result from the large-scale deployment of CAVs [4]. These challenges have recently prompted the design of communication systems that go beyond Level A, focusing on the content of the transmitted message and addressing the semantic and effectiveness communication problems formulated by Shannon and Weaver at Level B and Level C, respectively: *"how precisely do the transmitted symbols convey the desired meaning?"* (the semantic problem, Level B), and *"how effectively does the received meaning affect conduct in the desired way?"* (the effectiveness problem, Level C) [3].

Level B semantic communication systems have gained considerable attention in recent years. They focus on the extraction and transmission of the essential information needed to convey the meaning contained in specific data types such as text [5], audio [6], image [7], and video [8]. Most research in this area has focused on the Physical layer (PHY) layer, aiming to reduce the size of transmitted messages beyond the limits of conventional lossless and lossy compression algorithms [9] by

The work of Luca Lusvarghi was supported by the European Union under the 2023 MSCA Postdoctoral Fellowship program (project no. 101153845). This work was supported in part by MICIU/AEI/10.13039/501100011033 and "ERFD/EU" (project PID2023-150308OB-I00), and Generalitat Valenciana. *(Corresponding author: Luca Lusvarghi).*

Luca Lusvarghi, Javier Gozalvez, Baldomero Coll-Perales, and Miguel Sepulcre are with the Networked Systems Lab, Universidad Miguel Hernandez de Elche, Elche, 03202, Spain. E-mail: {llusvarghi, j.gozalvez, bcoll, msepulcre}@umh.es. Mohammad Irfan Khan, Seyhan Ucar, and Onur Altintas are with Toyota Motor North America R&D, InfoTech Labs, Mountain View, CA, 94043, USA. E-mail: {mohammad.irfan.khan, seyhan.ucar, onur.altintas}@toyota.com.

focusing on the exchange of meaning rather than mere data. Semantic communication systems typically require sophisticated AI-based techniques with software, hardware, and energy demands that may challenge their deployment on resource-constrained connected devices.

Recent contributions attempting to address the Level C problem have focused on curating the content of the transmitted messages by optimizing metrics such as the Age of Information (AoI) [10], Age of Incorrect Information (AoII) [11], and Value of Information (VoI) [12]. These proposals rely on the assumption that these metrics can help identify the most important information for the intended receivers. However, metrics like AoI, AoII, and VoI are context-agnostic combinations of objective information attributes (e.g., freshness, quality, precision), while context is key to understand the subjective meaning of the shared information and determine its impact on the intended receivers' behavior. In general, context refers to the circumstances or conditions under which a piece of information is exchanged and processed at the receiver [13]. It includes information about the receiver's characteristics and their evolution, the communication medium, and the physical environment where the exchange of information takes place. Understanding the intended receivers' context is a complex task that typically requires processing large amounts of rich and fresh contextual information. In most existing communication systems, connected devices do not typically share such contextual information. Consequently, context often has to be reconstructed individually (and not cooperatively) by the transmitter [14], leading to partial and potentially inaccurate estimates.

This paper proposes a novel semantic and task-oriented[1] communication paradigm that jointly addresses both Level B (semantic) and Level C (effectiveness) problems by making the search for relevance the core principle of the data exchange. As highlighted by linguists D. Wilson and D. Sperber in their relevance theory [15], the search for relevance is a key feature of human communication that allows humans to minimize the effort required to convey a desired meaning, i.e., maximize communication efficiency. In the relevance theory, relevance is a non-binary concept that captures the impact of a piece of information on the listener's perception of the world by intertwining its intrinsic meaning with the specific context in which it is exchanged. The proposed semantic and task-oriented communication paradigm brings the relevance-based principles of human communication to the domain of machine-type communications. In semantic and task-oriented communications, the transmitting devices curate the content of the transmitted messages based on its relevance for the intended receivers. The concept of relevance directly stems from the definition provided by the relevance theory, but is extended in our proposal to address both Level B and Level C aspects. We define the relevance of a piece of information $x$ as the context-dependent impact that the meaning (Level B) of $x$ has on the receiver's digital representation of the world and, ultimately, on the receiver's task (Level C). Like in the relevance theory, the relevance of a piece of information is intrinsically related with the intended receiver's context. Relevance's context-dependent nature allows it to more accurately capture the meaning and impact of the shared information at the intended receivers with respect to existing metrics (AoI, AoII, and VoI).

We believe that the V2X domain provides a unique ecosystem for developing and deploying the proposed semantic and task-oriented communication paradigm. First, CAVs are semantic-native devices. They are natively equipped with powerful processors (multiple high-end CPUs and GPUs) and AI algorithms capable to semantically process data collected by onboard sensors and extract the most relevant information (e.g., lane margins, positions of surrounding objects) needed for autonomous driving. Second, V2X communications inherently provide the rich contextual information required to achieve a correct context estimation, as CAVs continuously share their own data (e.g., their position, speed, and intended trajectory) as well as sensed data about the driving environment (e.g., the position of locally detected objects).

We qualitatively illustrate the operation and the potential of semantic and task-oriented V2X communications focusing on the cooperative perception use case [16]. In cooperative perception, vehicles share information about locally detected objects to collectively overcome their onboard sensors line-of-sight and range limitations and enhance their situational awareness. Several studies have shown that cooperative perception can challenge the scalability of V2X networks [17][18][19]. In the qualitative analysis, we show how, by transmitting only the most relevant information, each vehicle can reduce the amount of information per transmitted message without compromising the situational awareness of its intended receivers. This is key to improve the communication efficiency and, ultimately, the network's scalability. We corroborate these qualitative insights through a quantitative analysis that compares the performance of semantic and task-oriented V2X communications with existing V2X solutions, considering both unicast and broadcast communication scenarios. Our analysis shows that, by focusing on relevance, semantic and task-oriented V2X communications can increase the amount of relevant information transmitted by each vehicle to its intended receivers while preventing the transmission of unnecessary information. The obtained results show that semantic and task-oriented V2X communications can substantially reduce the amount of resources consumed by each vehicle and achieve a two-fold improvement in communication efficiency.

The remainder of this paper is organized as follows. Section II reviews related work on Level B and Level C semantic and task-oriented communications, with a particular focus on contributions in the V2X domain. Section III qualitatively illustrates the potential of semantic and task-oriented V2X communications within the context of cooperative perception. Section IV introduces the framework used for the design and quantitative analysis of semantic and task-oriented V2X communications while Section V describes its step-by-step operation. Section VI presents the numerical results and discusses practical deployment challenges. Finally, Section VII summarizes the main conclusions and contributions of this paper.

[1]The terms task-oriented and goal-oriented communications are interchangeably used in the literature.

## II. Related Work

The design of existing communication systems – including V2X – has traditionally been focused at Level A, and the majority of research efforts has concentrated on the design of novel techniques to guarantee the timely and reliable exchange of information with varying latency and reliability requirements. In V2X, Level A solutions include the design of robust and flexible MAC and PHY layers and message generation rules that control the rate and size of the generated messages. However, the scalability challenges introduced by the increasing data demand and number of connected devices – including the large-scale deployment of CAVs – have recently pushed the design of communication systems beyond Level A, towards Level B and Level C.

Most contributions in Level B semantic communications operate at the PHY layer, primarily focusing on Joint Source-Channel Coding (JSCC). These proposals generally target the design of semantic compression techniques that can extract and transmit the semantic information embedded in specific data types such as text, audio, image, and video, for example. In this context, semantic information refers to the essential set of features needed to convey the desired meaning of the exchanged information. For example, [5] introduces an end-to-end Transformer-based semantic communication system which can learn, extract, and transmit the essential semantic features of written text. A seminal contribution to semantic audio compression is presented in [6], where the authors extend their original work from [5] to introduce an end-to-end Deep Learning (DL)-based semantic communication system that employs 2D Convolutional Neural Networks (CNNs). The authors show that this system can effectively learn, extract, and transmit the semantic information of raw speech signals. Contributions in [7] and [8] address semantic compression for images and video. In [7], the authors present a DL-enabled semantic communication system featuring two matched CNNs connected within an end-to-end JSCC architecture, where the transmitter extracts and maps the critical spatial features of an image to a complex channel codeword. In [8], the authors address the design of a semantic video conferencing architecture using an end-to-end DL network that combines CNNs with hourglass networks to extract and transmit only crucial keypoints of video frames.

In the V2X domain, contributions in Level B semantic communications primarily focus on developing semantic compression techniques for image transmission, aiming to reduce the volume of information that vehicles exchange. For example, [9] compares the performance of semantic segmentation with lossy compression techniques, demonstrating that transmitting semantic information extracted from raw images can reduce network load, decrease end-to-end latency, and lower energy consumption. The contributions in [20] and [21] also make use of semantic segmentation. In [20], the authors design an autoencoder-based lossy compression technique that focuses image compression on regions of interest associated with semantically important classes, such as vehicles and trucks. The contribution in [21] presents an end-to-end Image Segmentation Semantic Communication (ISSC) system, designed to exchange only the essential semantic features present in the input image. In contrast, the work in [22] does not explicitly employ semantic segmentation. Instead, it presents a DL-based cooperative semantic communication system in which multiple transmitting vehicles collaboratively extract and share semantic information extracted from images.

Current proposals to address the effectiveness problem at Level C leverage the concepts of AoI, AoII, and VoI for designing task-oriented communication systems. AoI captures the information freshness considering both the end-to-end transmission latency and the rate at which data is generated and received. In AoI-based systems, the content of transmitted messages is curated to minimize the AoI at the receiver, operating under the assumption that fresher information is more valuable for the receiver's task [10]. AoII extends the AoI concept by combining the freshness of information and its informativeness [11]. A piece of information is considered informative when it is not already known to the receiver. AoII-based systems assume that fresh and informative information is more valuable for the receiver's task and prioritize its transmission.

Existing VoI definitions are highly task-dependent and encompass a broader range of attributes with respect to the AoI and AoII metrics. These attributes include, for example, information quality, precision, proximity, redundancy, freshness and informativeness. In the V2X domain, VoI-based communication systems use the VoI to determine what information vehicles should share. The study in [23] defines VoI as a weighted combination of proximity, timeliness, and quality of the information, with the specific weights adjusted according to the V2X use case. In [24], VoI for a remote driving task is defined as the reduction in path tracking error experienced by the receiving vehicle after correctly decoding a transmitted message. The study in [18] proposes a VoI definition based on relative entropy in the context of cooperative perception to decide which detected objects should be included in cooperative perception messages. Similarly, [19] introduces a VoI definition based on object dynamics, applying it to cooperative perception to transmit only the most dynamic detected objects, i.e., those with the greatest changes in speed, position, or heading. This approach is based on the premise that accurate perception of the driving environment requires more frequent updates on highly dynamic objects. Additional VoI definitions based on detected objects' redundancy, perception quality, and classification confidence have recently been included in the cooperative perception standard [16] to regulate the amount of information shared by vehicles.

By curating the information shared by vehicles, existing Level C contributions in the V2X domain have made important advances in terms of communication efficiency. However, the relevance of information is highly context-dependent, and the limitations of current Level C metrics become evident when considering cooperative perception, for example. A fresh (AoI [10]) and informative (AoII [11]) update about an unseen (entropy-based VoI [18]) and high-speed (dynamics-based VoI [19]) detected object may not be relevant to a receiving vehicle if the object is, for example, traveling in the opposite direction and has already passed the receiving vehicle. The importance to capture the context-dependence relevance of information is a central aspect of the proposed semantic and task-oriented V2X communication paradigm.

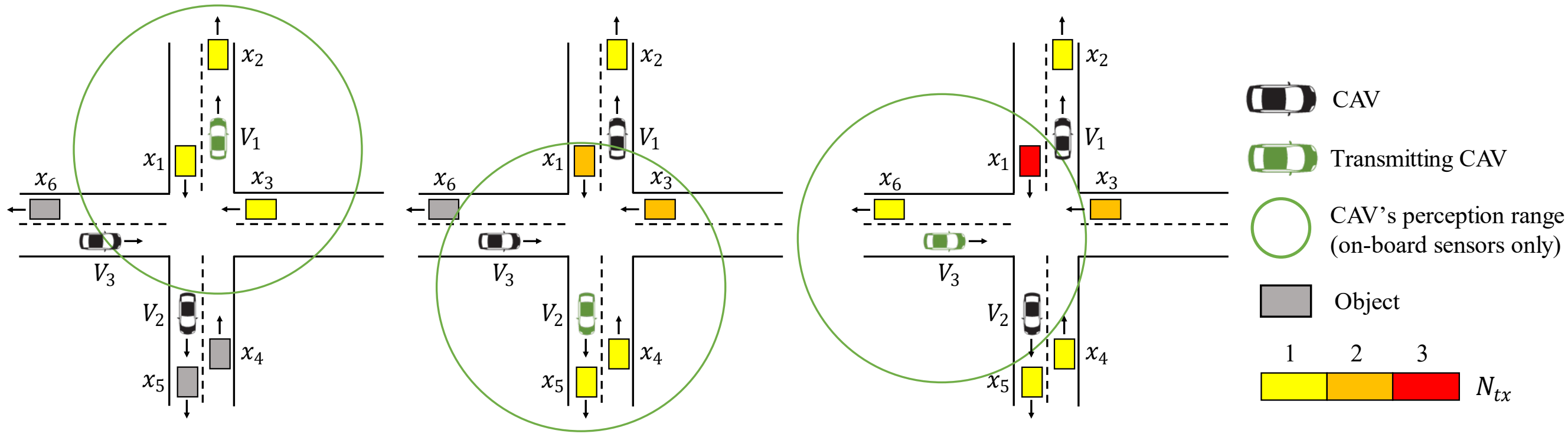

(a) Transmitting CAV: $V_1$. (b) Transmitting CAV: $V_2$. (c) Transmitting CAV: $V_3$.

**Fig. 1.** Cooperative perception.

## III. Semantic and Task-Oriented V2X Communications: An Illustrative Example with Cooperative Perception

In the V2X domain, relevance can be defined as the context-dependent impact that the meaning of a piece of information has on the receiving vehicle's understanding of the driving environment and, ultimately, on its driving. Relevance in the V2X domain is highly dependent on the receiver's context, which is defined by a dynamic and complex set of spatio-temporal interrelationships between contextual information elements like:

- Road topology, driving rules, and any other road information describing the driving environment.
- The current position, speed, and heading of the intended receiver. More generally, any information about the current and future state of the intended receiver, including its planned trajectory.
- Previously exchanged messages containing information about the intended receiver's surrounding environment or any information about the data already available at the intended receiver.
- Information about the V2X communication link, including the channel load and the transmitter-receiver distance.

Relevance's context-dependent nature is particularly evident in cooperative perception (also known as collective perception or sensor data sharing) [16]. In cooperative perception, contextualizing the meaning and attributes of locally detected objects is essential to accurately determine their relevance for a receiving vehicle and prevent the transmission of unnecessary (or irrelevant) information. For example, consider a transmitting vehicle that detects a pedestrian (object) crossing an intersection using its onboard sensors. If the receiving vehicle is approaching the same intersection (context), the information about the pedestrian represents a safety-critical situation (meaning) that should be avoided, and its relevance is high, provided the information is accurate and timely (attributes). Conversely, if the receiving vehicle is leaving the intersection, the presence of the pedestrian does not represent a safety risk for it, and the relevance of the information about the pedestrian is low, regardless of its attributes.

We illustrate the key principles and the potential of semantic and task-oriented V2X communications considering a cooperative perception example. Fig. 1 illustrates the standard-compliant operation of cooperative perception (i.e., without semantic and task-oriented V2X communications) in an intersection scenario with three communicating vehicles ($V_1$, $V_2$, and $V_3$) and six objects ($x_k$, $k = 1,\dots,6$). In this scenario[2], we assume that all three vehicles are within V2X communication range and that there are no packet collisions or transmission errors. In Fig. 1(a), $V_1$ shares information with $V_2$ and $V_3$ about all three objects it has detected within its perception range ($x_1$, $x_2$, and $x_3$). In Fig. 1(b), $V_2$ reports to $V_1$ and $V_3$ about all four objects it has detected ($x_1$, $x_3$, $x_4$ and $x_5$). Similarly, in Fig. 1(c), $V_3$ informs $V_1$ and $V_2$ about the two objects it has detected ($x_1$ and $x_6$). Thanks to cooperative perception, all vehicles ($V_1$, $V_2$ and $V_3$) receive information about all six objects in the scenario, even those outside the perception range of their onboard sensors. However, Fig. 1 also highlights some of the challenges and inefficiencies associated with cooperative perception, such as redundancy. For example, all three vehicles transmit information about $x_1$. As a result, $x_1$ is redundantly transmitted a total of $N_{tx} = 3$ times in Fig. 1(c). Similarly, information about $x_3$ is transmitted a total of $N_{tx} = 2$ times (by both $V_1$ and $V_2$). While some redundancy can be useful for confirming or cross-verifying local data, excessive redundancy can unnecessarily consume limited spectrum resources and compromise the scalability of V2X networks. Overloading the V2X network with redundant information may delay or block the transmission of relevant data, ultimately impacting the effectiveness of cooperative perception [17]. The example in Fig. 1 also highlights that cooperative perception can result in the transmission of non-redundant yet irrelevant information. For instance, object $x_2$, detected and transmitted by $V_1$ in Fig. 1(a), has low relevance for $V_2$ and $V_3$ since $x_2$ is leaving the intersection and does not influence $V_2$ (driving in the opposite direction) or $V_3$ (approaching the intersection). Similarly, the object $x_5$ transmitted by $V_2$ in Fig. 1(b) and the object $x_6$ transmitted by $V_3$ in Fig. 1(c) have minimal relevance for the respective receiving vehicles.

[2] For simplicity, we assume that the detected objects reported by CAVs consists of only non-connected vehicles. In standardized cooperative perception, CAVs report different types of non-connected objects (vehicles and other road users) and also other CAVs within their perception range.

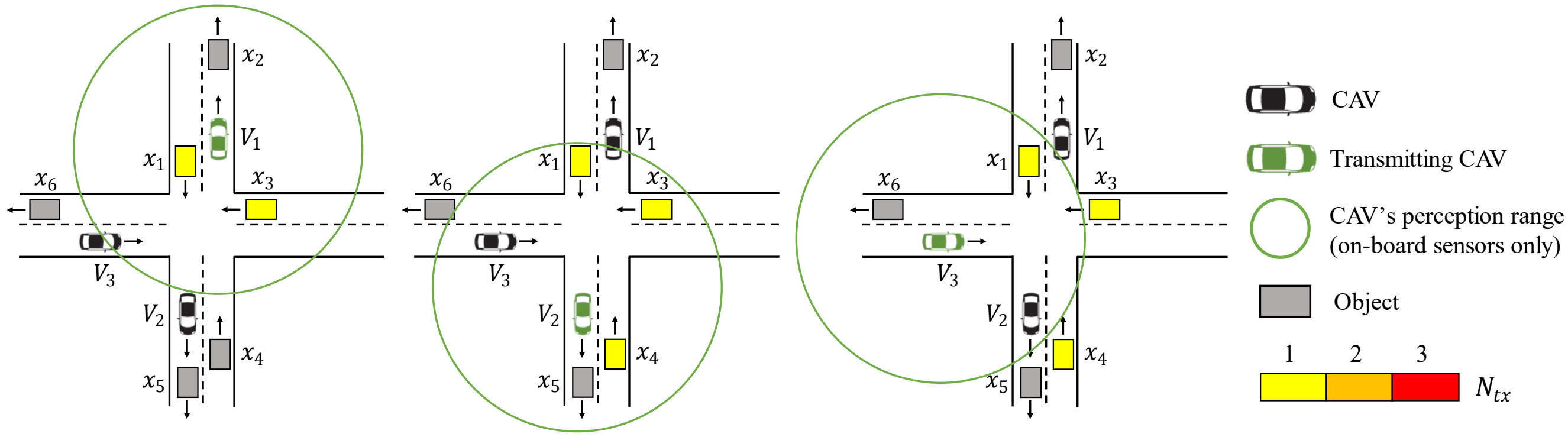

(a) Transmitting CAV: $V_1$ (b) Transmitting CAV: $V_2$ (c) Transmitting CAV: $V_3$

**Fig. 2.** Cooperative perception with semantic and task-oriented V2X communications.

Standard-compliant cooperative perception relies on the frequent transmission of locally detected objects without considering their relevance to the receiving vehicles. Fig. 2 illustrates how semantic and task-oriented V2X communications would impact cooperative perception by filtering the transmitted objects based on their context-dependent relevance to the intended receivers. In Fig. 2, each vehicle only transmits the objects that it deems relevant to the receiving vehicles. An object is considered relevant to an intended receiver if it can influence the receiver's driving decisions. Vice versa, an object is considered irrelevant – and, therefore, excluded from the transmitted message – if it does not have any impact on the receiver's driving or it has already been transmitted by another vehicle, i.e., it is redundant. For example, in Fig. 2(a), $V_1$ transmits only objects $x_1$ and $x_3$, as they are approaching the intersection and may impact the driving decisions of $V_3$, which is also approaching the intersection. Although $V_1$ detects also $x_2$, it does not report it, as it estimates that it is not relevant for the intended receivers ($V_2$ and $V_3$) since $x_2$ is exiting the intersection and does not impact their mobility. Similarly, in Fig. 2(b), $V_2$ detects objects $x_1$, $x_3$, $x_4$ and $x_5$, but only transmits object $x_4$. $V_2$ does not transmit objects $x_1$ and $x_3$ because they were already reported by $V_1$ and, therefore, would be redundant. Moreover, $V_2$ does not transmit $x_5$ since it is leaving the intersection and has no relevance for $V_3$ (approaching the intersection) or $V_1$ (moving in the opposite direction), even if $x_5$ is not redundant. Similarly, $V_3$ does not transmit object $x_6$ in Fig. 2(c), as it estimates that it is irrelevant for both $V_1$ and $V_2$ due to their driving directions. $V_3$ does not transmit object $x_1$ as it estimates its information to be redundant for both $V_1$ and $V_2$ since $V_1$ already reported about $x_1$ in Fig. 2(a).

The comparison between Fig. 1 and Fig. 2 qualitatively highlights the potential advantages of semantic and task-oriented V2X communications in cooperative perception. In Fig. 1, vehicles transmit all locally detected objects regardless of their relevance for the intended receivers, with some information being transmitted multiple times ($N_{tx} > 1$). In contrast, Fig. 2 shows that semantic and task-oriented V2X communications curate the content of the transmitted messages based on its context-dependent relevance. With semantic and task-oriented V2X communications, vehicles can reduce the amount of transmitted data while ensuring that relevant situational awareness information is shared. By focusing on relevance, semantic and task-oriented V2X communications can enhance the scalability of V2X networks and guarantee the large-scale effectiveness of the supported use cases. We should note that the potential advantages of semantic and task-oriented V2X communications are not confined to the cooperative perception context, and can be extended to other use cases such as cooperative awareness and cooperative driving.

## IV. Framework

We consider a driving scenario of $W$ x $H$ m$^2$ with $N$ CAVs, each denoted by $V_n$, $n = 1, \dots, N$. Without loss of generality, each vehicle $V_n$ is randomly assigned a unique pair of origin-destination driving points within the scenario. We assume that for each origin-destination pair, there is a single path $p_n$ that each vehicle can follow to reach its destination. The internal state of each vehicle $V_n$ is denoted as $s_n$ and we define $s_n$ as a combination of endogenous and exogenous variables. Endogenous variables ($s_n^{end}$) represent the intrinsic characteristics of the vehicle, such as its perception range, and all driving aspects directly controlled by the vehicle, including its position, speed, acceleration, destination and path $p_n$. Exogenous variables ($s_n^{ex}$) represent information that is external to the vehicle. In the context of cooperative perception, exogenous variables represent objects populating the driving environment (e.g., non-connected vehicles, vulnerable road users, obstacles); these objects are characterized by their position, size, or speed, for example. We define the set of exogenous variables available at $V_n$ as:

$$s_n^{ex} = s_{n,loc}^{ex} \cup s_{n,V2X}^{ex}, \tag{1}$$

where $s_{n,loc}^{ex}$ represents the subset of exogenous variables collected locally by $V_n$ using its onboard sensors, and $s_{n,V2X}^{ex}$ represents the subset of exogenous variables received by $V_n$ from other vehicles via V2X communications.

Let $\mathcal{K}$ denote the set of $K$ exogenous variables available in the driving scenario, and let $x_k$, with $k = 1, \dots, K$, represent the $k$-th exogenous variable. We model the distribution of exogenous variables across the driving scenario using a 2D Poisson Point Process (PPP). We denote with $P(x_k \in s_{n,loc}^{ex})$ the probability that the exogenous variable $x_k$ can be locally detected by $V_n$ using its onboard sensors. In a cooperative perception context, $P(x_k \in s_{n,loc}^{ex})$ represents the detection probability, i.e., the probability that an object $x_k$ is detected by $V_n$ using its onboard sensors. This probability is modeled as a logistic function of $D_{k,n}$, the distance between $x_k$ and $V_n$:

$$P\left(x_k \in s_{n,loc}^{ex}\right) = \frac{1}{1 + a_1 e^{-a_2 (D_{k,n} - a_3)}}, \quad (2)$$

where $a_1$, $a_2$, and $a_3$ are fixed values.

Vehicles exchange the exogenous variables locally detected using onboard sensors on a time-slotted V2X communication system. We assume that there are no transmission errors or packet collisions. Vehicles take turns accessing the channel and, in each time slot, each vehicle can be in one of two possible states: Transmit or Receive. In the Transmit state, the transmitting vehicle ($V_T$) first updates its set of locally collected exogenous variables ($s_{T,loc}^{ex}$) based on the detection probability $P\left(x_k \in s_{T,loc}^{ex}\right)$ defined in (2). Then, $V_T$ generates a V2X message that includes its own position and speed (i.e., its endogenous variables $s_T^{end}$) and a selected subset $s_{T,tx}^{ex}$ of the locally collected exogenous variables (where $s_{T,tx}^{ex} \subset s_{T,loc}^{ex}$). In the V2X domain, vehicles are allowed to transmit only local information to avoid the propagation of detection errors on the V2X network. The number of exogenous variables included in each message ($\left|s_{T,tx}^{ex}\right|$) must comply with some communication constraint $\Gamma$. $\Gamma$ represents the maximum number of exogenous variables that each vehicle can transmit in a message. The number of transmitted exogenous variables is a proxy of the message size and $\Gamma$ represents the network-level communication constraint that can be enforced by V2X congestion control mechanisms to keep the V2X network load below pre-defined thresholds. If $\left|s_{T,tx}^{ex}\right|$ exceeds $\Gamma$, the subset of exogenous variables included in the generated message is filtered until the $\Gamma$ constraint is met. In the Receive state, a receiving vehicle ($V_R$) decodes the message transmitted by $V_T$ and updates the subset of exogenous variables received via V2X communications ($s_{R,V2X}^{ex}$) with the received information, i.e., $s_{T,tx}^{ex}$. We assume that the information received from a transmitting vehicle expires after $N$ slots, meaning it is valid only until the same transmitting vehicle transmits a new set of locally collected exogeneous variables in the next communication cycle. At each vehicle, a communication cycle is defined as the time interval between two consecutive transmission attempts.

## V. Semantic and Task-Oriented V2X Communications

With semantic and task-oriented V2X communications, vehicles transmit only the subset of locally collected exogenous variables that they estimate to be relevant to their intended receivers. We measure the context-dependent relevance of an exogenous variable $x_k$ for a vehicle $V_n$ with its semantic value $w_{n,k}$. We define a contextual relevance function $f_n(\mathcal{K})$ to represent the relevance of any exogenous variable $x_k$ in the driving environment for each vehicle $V_n$. Without loss of generality, the contextual relevance function used in this study is a discrete weighting function that assigns non-uniform semantic values $w_{n,k}$ to each exogenous variable $x_k \in \mathcal{K}$. Exogenous variables are randomly assigned a low or high relevance depending on their distance from $V_n$ and the vehicle's speed and planned path ($p_n$). Exogenous variables that are closer to $V_n$ are more likely to have a high relevance. Exogenous variables with low relevance have a negligible semantic value, i.e., $w_{n,k} \cong 0$, and the fraction of low relevance variables with respect to the total number of exogenous variables in the driving environment ($K$) is denoted by $\Delta_L$. Low relevance exogenous variables are also referred to as irrelevant exogenous variables.

An object or exogenous variable may have high or low relevance (semantic value) for a given vehicle, and the same object can be highly relevant for one vehicle while being irrelevant for another, depending on their respective contexts. The contextual relevance function $f_n(\mathcal{K})$ is designed to provide a simplified yet realistic representation of the context-dependent relevance that objects (exogenous variables) in the driving environment have for different vehicles. This is illustrated in Fig. 3, which shows the contextual relevance functions $f_3(\mathcal{K})$ and $f_2(\mathcal{K})$ of vehicles $V_3$ and $V_2$ in the cooperative perception example depicted in Fig. 1 and Fig. 2. Since both $V_3$ and object $x_1$ are approaching the intersection, the presence of $x_1$ may influence the driving behaviour of $V_3$. Accordingly, $x_1$ is assigned a high semantic value $w_{3,1}$ in the contextual relevance function of $V_3$ (see Fig. 3(a)), indicating that $x_1$ is highly relevant to $V_3$. For the same reason, objects $x_3$ and $x_4$ are also assigned high semantic values ($w_{3,3}$ and $w_{3,4}$) in $V_3$'s contextual relevance function. On the other hand, objects $x_2$, $x_5$, and $x_6$ are leaving the intersection and therefore do not affect $V_3$'s driving decisions. This is reflected in the low semantic values $w_{3,2}, w_{3,5}, w_{3,6}$ assigned to these objects in $V_3$'s contextual relevance function (see Fig. 3(a)). We employ a randomization factor $p$ to account for the probability that the contextual relevance functions of two neighboring vehicles are uncorrelated and that, therefore, a detected object has completely different relevance for them. In the cooperative perception example shown in Fig. 1 and Fig. 2, this is the case of object $x_1$ with respect to vehicles $V_3$ and $V_2$. Despite their close proximity, $x_1$ is highly relevant for $V_3$ (which is also approaching the intersection) while it is irrelevant for $V_2$ (which is leaving the intersection), as indicated by the low semantic value $w_{2,1}$ assigned by $f_2(\mathcal{K})$ in Fig. 3(b). When the contextual relevance functions of two neighbouring vehicles are correlated (with probability $1 - p$), we employ a distance-based correlation coefficient to model the probability that a detected object (i.e., an exogenous variable) has the same relevance for both vehicles. In Fig. 1 and Fig. 2, this would be the case of object $x_5$ with respect to vehicles $V_2$ and $V_3$ if, for example, $V_3$ decides to turn right at the intersection. In such a scenario, object $x_5$ would block the lane of both $V_2$ and $V_3$, thereby limiting their speed.

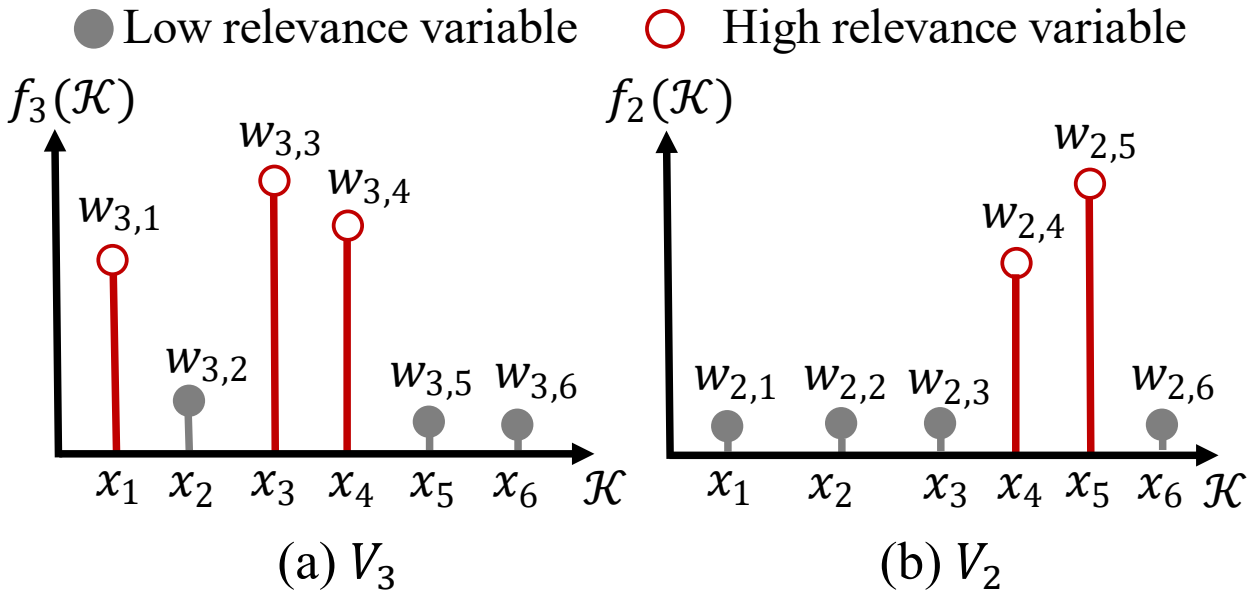

(a) $V_3$ (b) $V_2$

**Fig. 3.** Contextual relevance function $f_n(\mathcal{K})$ design.

The operation of semantic and task-oriented V2X communications is illustrated in Fig. 4. A transmitting vehicle first estimates which locally collected exogenous variables ($s^{ex}_{T,loc}$) are redundant for its intended receivers (*redundancy estimation* in Fig. 4). To this aim, it checks the content of messages previously exchanged over the V2X network, assuming that the content of any correctly decoded message is also available, i.e., redundant, at all its intended receivers. This is how redundancy estimation is typically performed in cooperative perception [16]. The transmitting vehicle does not transmit redundant exogeneous variables as it estimates that transmitting again these variables would not improve the intended receiver's understanding of the driving environment, making them irrelevant.

Then, the transmitting vehicle processes the available contextual information to estimate the context-dependent relevance of non-redundant locally collected exogenous variables for all its intended receivers (*relevance estimation* in Fig. 4). We denote with $\mathcal{R}$ the set of all intended receivers. Contextual information consists of (i) all the exogenous variables exchanged over the V2X network and (ii) the position, speed, and planned path of each intended receiver (i.e., its endogenous variables). The transmitting vehicle uses a relevance model to estimate the relevance (or semantic value) $\widehat{w}_{R,k}$ of each non-redundant locally collected exogenous variable $x_k$ ($x_k \in s^{ex}_{T,loc}$) for each intended receiver $V_R$. The relevance model leverages the available contextual information to estimate the contextual relevance function $\hat{f}_R(\mathcal{K})$ of each intended receiver $V_R$ ($R \in \mathcal{R}$). To capture realistic conditions, we assume that the estimation of the semantic value $\widehat{w}_{R,k}$ assigned by $\hat{f}_R(\mathcal{K})$ to a locally collected exogenous variable $x_k$ is subject to an estimation error $\varepsilon$. We model the estimated semantic value $\widehat{w}_{R,k}$ as a uniform random variable distributed within an estimation interval centred on the actual semantic value ($w_{R,k}$) and with a width that depends on $\varepsilon$. We define the estimation error $\varepsilon$ using a logistic function:

$$\varepsilon = \frac{1}{1 + a_4 e^{-a_5(|s^{ex}_T| - a_6)}}, \tag{3}$$

where $a_4$, $a_5$, and $a_6$ are fixed values and $|s^{ex}_T|$ represents the amount of contextual information (exogenous variables) available at the transmitting vehicle. We assume that the transmitter can reduce the relevance model's estimation error $\varepsilon$ and improve the accuracy of the $\widehat{w}_{R,k}$ estimate as it collects more contextual information from the driving environment (i.e., as $|s^{ex}_T|$ increases).

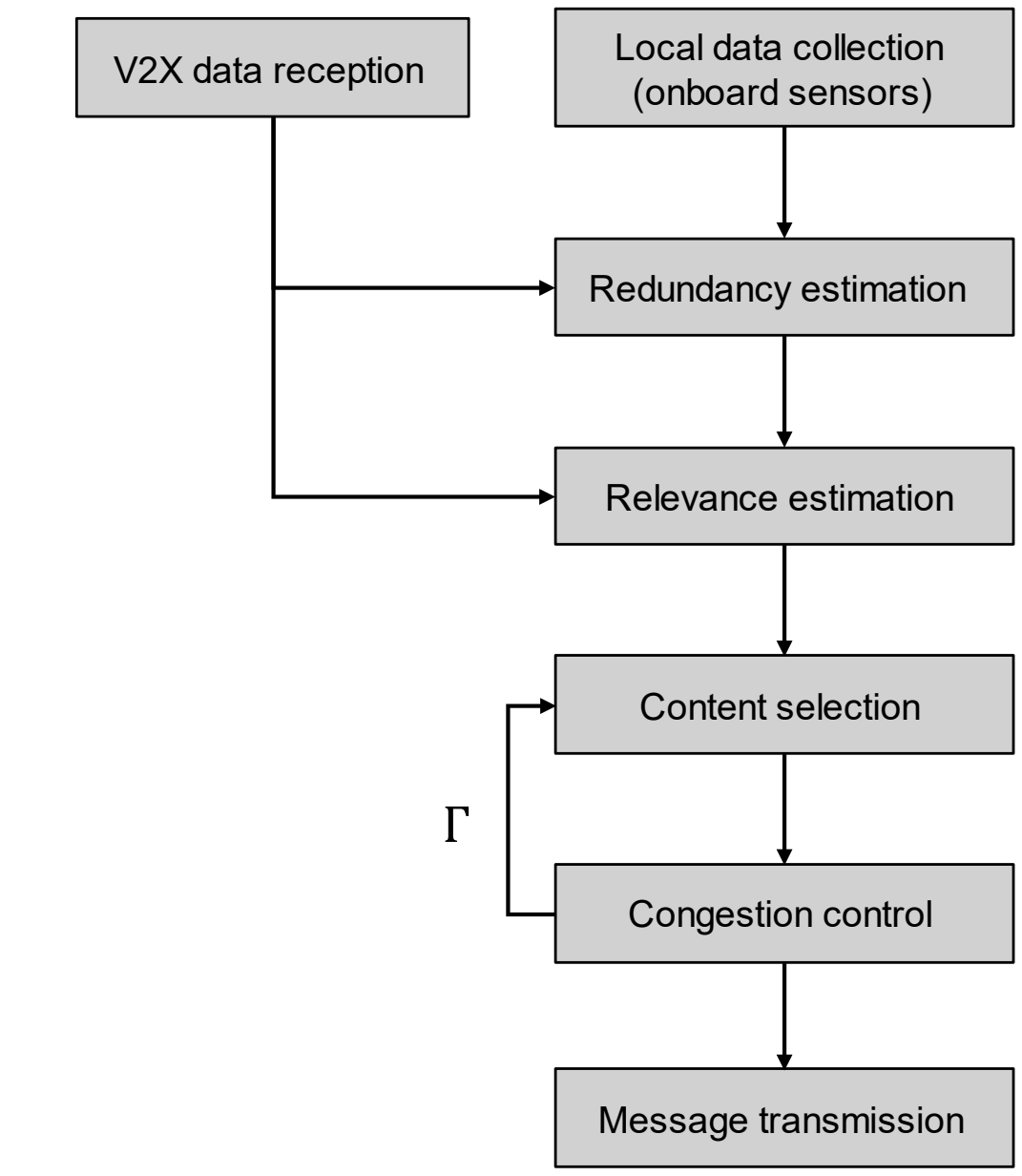

**Fig. 4.** Semantic and task-oriented V2X communications.

The transmitting vehicle leverages the estimated relevance (or semantic value) $\widehat{w}_{R,k}$ of each locally detected exogenous variable ($x_k \in s^{ex}_{T,loc}$) to select the content of the generated messages. Note that the relevance of each $x_k$ is estimated considering all intended receivers ($\mathcal{R}$) and, therefore, the relevance of each $x_k$ is a multi-dimensional estimate $\widehat{w}_{\mathcal{R},k}$ with one value per intended receiver:

$$\widehat{w}_{\mathcal{R},k} = \left[\widehat{w}_{R,k}\right]_{R \in \mathcal{R}}. \tag{4}$$

The content-selection process (*content selection* in Fig. 4) aims to maximize the relevance of the generated messages while respecting the communication constraint $\Gamma$ enforced by the congestion control mechanism and avoiding the transmission of irrelevant information. Congestion control establishes the value of $\Gamma$ that represents the maximum number of exogenous variables that each vehicle can transmit in a message. The operation of the content-selection process is as follows. The transmitting vehicle ranks exogenous variables based on their maximum relevance across all intended receivers $R \in \mathcal{R}$. The maximum relevance of an exogenous variable can be analytically expressed as $\max_{R \in \mathcal{R}} \widehat{w}_{\mathcal{R},k}$. Exogenous variables with a maximum relevance below $w_{min}$ are filtered out and not included in the generated message. $w_{min}$ is a semantic value threshold that distinguishes between relevant and irrelevant information. Among the remaining exogenous variables (with maximum relevance greater than $w_{min}$), the top $\Gamma$ ones are included in the generated message. The operation of the content-selection process can be expressed analytically as:

$$\begin{aligned} s^{ex}_{T,tx} &= \arg\max_{x_k \in s^{ex}_{T,loc}} \max_{R \in \mathcal{R}} \widehat{w}_{\mathcal{R},k} \\ \text{s.t. } &\max_{R \in \mathcal{R}} \widehat{w}_{\mathcal{R},k} > w_{min} \\ &\left|s^{ex}_{T,tx}\right| < \Gamma \end{aligned} \tag{5}$$

For the sake of brevity, we will refer to the semantic and task-oriented V2X communication scheme as the *Semantic* V2X communication scheme in the rest of this work.

## VI. Numerical Evaluation

This Section presents a numerical evaluation of the potential of semantic and task-oriented V2X communications to efficiently and scalably accomplish the communication goal: provide the intended receivers with the most relevant driving environment information. Unless otherwise specified, we consider a driving scenario of 800 x 200 m$^2$ with a total of $|K|$ = 110 exogenous variables. We configure the coefficients $a_1$, $a_2$, and $a_3$ of the detection probability in (2) to 0.08, -0.08, and 60, respectively, to obtain a local perception range equal to 150 m. The local perception range is defined as the distance at which the probability $P(x_k \in s_{n,loc}^{ex})$ of locally detecting an exogenous variable $x_k$ reaches zero. This setting is in line with the local perception range that characterizes the object detection probability of existing cooperative perception studies such as [17] and [19]. Vehicles are capable of communicating in either unicast or broadcast mode and, without loss of generality, we assume an ideal V2X communication channel, meaning that all transmitted messages are correctly decoded by the receivers. We consider that only 30% of the of the $|K|$ exogenous variables are highly relevant ($\Delta_L$= 0.7). Low-relevance variables are assigned a semantic value $w_{n,k}$ of 0, while the semantic value of high-relevance variables is randomly sampled from the range [0.5,1]. The contextual relevance functions of two neighbouring vehicles are uncorrelated with a probability $p$ equal to 0.5. For correlated contextual relevance functions, the correlation coefficient depends on the distance between the vehicles. The coefficient is set to 0.9 when the distance is less than 100 m, and linearly decreases to zero at a distance of 400 m. The coefficients $a_4$, $a_5$, and $a_6$ in (3) that define the relevance model's estimation error $\varepsilon$ are set to 1, -0.5, and 26.

### A. Benchmark V2X Communication Schemes

We compare the performance of the semantic and task-oriented V2X communications against four benchmark V2X communication schemes: *Ideal Semantic, Redundancy Mitigation (RM)*, *Inclusion Rate Control (IRC)*, and *Baseline*.

The *Ideal Semantic* scheme represents an ideal implementation of the *Semantic* V2X communication scheme where the redundancy and relevance of locally collected exogenous variables is perfectly estimated.

The *RM* scheme was originally proposed in the context of cooperative perception to avoid the transmission of redundant information and improve the scalability of V2X networks [17]. With *RM*, a vehicle includes in a message the subset of locally collected exogenous variables ($s_{T,loc}^{ex}$) that are not redundant following the redundancy estimation mechanism implemented for *Semantic* V2X communications. If the number of non-redundant exogeneous variables is larger than Γ, then a random selection of Γ non-redundant exogenous variables is included in the transmitted message to comply with the communication constraint imposed by congestion control. *RM* filters out redundant exogenous variables but it may still transmit non-redundant exogenous variables that are irrelevant to the intended receivers.

The *IRC* scheme is part of the cooperative perception standard [16]. With *IRC*, a vehicle initially attempts to include all locally collected exogenous variables in the generated message. If the number of variables in the generated message ($s_{T,tx}^{ex}$) is larger than the constraint Γ, the *IRC* scheme removes redundant variables until $|s_{T,tx}^{ex}| = \Gamma$, i.e., until the constraint is met. Redundant variables are estimated following the same redundancy estimation mechanism used in *Semantic* V2X and *RM*. If the number of exogenous variables in the generated message exceeds Γ after removing all redundant exogenous variables, a random selection of Γ non-redundant exogeneous variables is included in the transmitted message. The difference between *IRC* and *RM* lies in the fact that *IRC* may still transmit redundant variables if this is permitted by Γ, while *RM* filters out all redundant variables independently of Γ.

With the *Baseline* scheme, a vehicle attempts to include all locally collected exogenous variables ($s_{T,loc}^{ex}$) in the generated messages ($s_{T,tx}^{ex} = s_{T,loc}^{ex}$), like in Fig. 1. If the number of locally detected exogenous variables exceeds Γ, the number of exogenous variables included in the message ($s_{T,tx}^{ex}$) must be reduced to meet the limit imposed by Γ. In this case, the *Baseline* scheme randomly selects Γ variables from all locally detected exogeneous variables ($s_{T,loc}^{ex}$) without making any distinction between redundant and non-redundant variables.

### B. Unicast

We first consider a unicast scenario with $N$ = 2 communicating CAVs. Fig. 5 compares the capability of the five V2X schemes to achieve the communication goal, namely, conveying the relevant information to the intended receiver. The figure reports the High-Relevance variables Ratio (HRR) as a function of the communication constraint Γ. The HRR measures the ratio between the number of high-relevance exogenous variables available at each vehicle and the total number of high-relevance exogenous variables $|K|$ available in the driving scenario. The considered Γ values range from an extremely resource-constrained communication scenario (Γ = 1) to an unconstrained scenario (Γ = 25 in our scenario) where the number of exogenous variables locally detected by each vehicle is always lower than Γ and CAVs can transmit them all. Fig. 5 shows that the *Ideal Semantic* and *Semantic* V2X schemes provide the intended receiver with a larger fraction of high-relevance exogenous variables compared to the *Baseline, IRC,* and *RM* schemes. Existing V2X schemes (*Baseline, IRC,* and *RM*) can only achieve HRR values similar to the semantic schemes when the communication constraints Γ are relaxed, i.e., when the communication bandwidth is not restricted and vehicles can transmit information about a larger set of exogeneous variables. In contrast, the *Ideal Semantic* and *Semantic* V2X schemes can provide a large fraction of high-relevance exogeneous variables even when Γ is low. These results underscore the potential of semantic and task-oriented V2X communications to improve the communication efficiency and convey the most relevant information to the intended receiver also in resource-constrained scenarios. The comparison between the *Semantic* and *Ideal Semantic* shows that the benefits of semantic and task-oriented V2X communications depend on the vehicles' capability to accurately estimate the contextual relevance function of their intended receiver. Nevertheless, Fig. 5 shows that semantic and task-oriented V2X communications significantly outperforms existing V2X schemes (*Baseline, IRC,* and *RM*), even with imperfect estimations (*Semantic*).

In addition, our analysis sheds light on the impact that the filtering of different types of information has on the vehicles' capability to maximize the HRR performance and accomplish the communication goal. The comparison between *Baseline* and *RM* (or *IRC*) schemes in Fig. 5 shows that the filtering of redundant information has negligible impact on the HRR performance. In Fig. 5, the *RM, IRC,* and *Baseline* V2X schemes exhibit similar HRR levels across all values of Γ. These results indicate that simply avoiding (*RM*) or reducing (*IRC*) the transmission of redundant information does not necessarily ensure the reception of the most relevant information and that, therefore, redundancy mitigation schemes alone can only partially address the scalability challenges of future V2X networks. In contrast, the comparison between *RM* (or *IRC*) and *Semantic* (or *Ideal Semantic*) schemes shows that the filtering of both redundant and irrelevant information is the key to significantly increase the amount of relevant information delivered to the intended receivers.

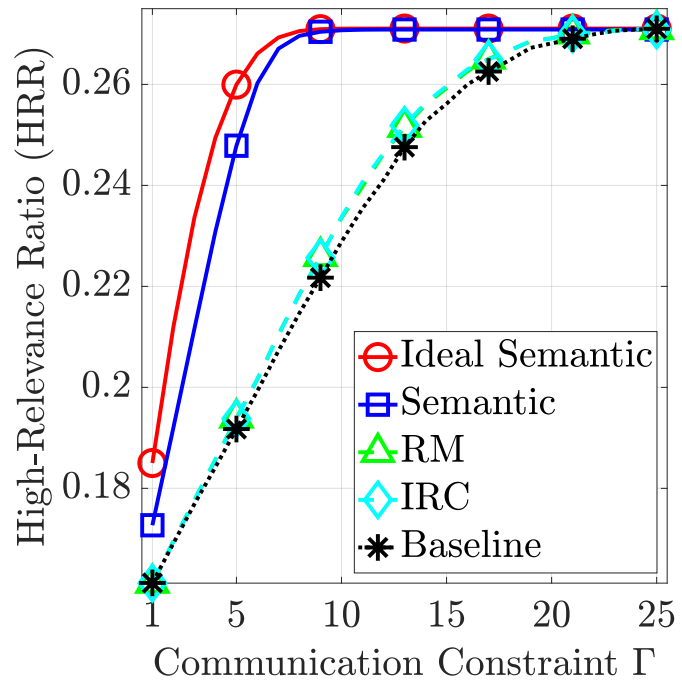

**Fig. 5.** HRR. Unicast scenario.

The capability of the V2X schemes to accomplish the communication goal and transmit the most relevant information to the intended receiver (Fig. 5) directly stems from their capability to semantically curate the content of transmitted messages. Fig. 6(a) presents the average semantic value ($\overline{SV}$) of the received messages when employing the five V2X schemes. At each receiving vehicle $V_R$, the semantic value (SV) of a transmitted message corresponds to the sum of the relevance (or semantic value) $w_{R,k}$ of each exogenous variable $x_k \in s^{ex}_{T,tx}$ included in it. The $\overline{SV}$ of a transmitted message is obtained by averaging its semantic value measured across all receiving vehicles. Note that the semantic value of an exogenous variable $x_k$ is zero if it is already available at the intended receiver (i.e., it is redundant), as we assume that redundant information does not improve the receiver's understanding of the world and hence, has no impact on its driving. Fig. 6(a) shows that the *Ideal Semantic* and *Semantic* schemes include more relevant information in their transmitted messages, achieving higher $\overline{SV}$ values than the *RM, IRC,* and *Baseline* schemes. The *Semantic* scheme achieves lower $\overline{SV}$ values than the *Ideal Semantic* as its estimate of the intended receiver's contextual relevance function is not entirely accurate. Reducing the relevance model's estimation error is key to unlock the full potential of semantic and task-oriented V2X communications. Additionally, Fig. 6(a) shows that relevance-agnostic schemes (*RM, IRC,* and *Baseline*) are particularly ineffective at ensuring the transmission of relevant content (i.e., with high semantic value) under low-medium Γ values or stringent communication constraints. These schemes do not filter exogeneous variables based on their relevance to the intended receiver and may attempt to transmit a number of variables $|s^{ex}_{T,tx}|$ that exceeds the communication constraint Γ. In this case, vehicles can transmit only Γ variables and these variables are randomly selected among the locally detected ones. The random filtering can inadvertently exclude exogeneous variables with high relevance or semantic value, as highlighted in Fig. 6(a). The importance of semantically curating the content of the transmitted messages decreases when Γ exceeds the average number of locally detected exogenous variables $|\overline{s^{ex}_{n,loc}}|$. The value of $|\overline{s^{ex}_{n,loc}}|$ is 15 in our scenario. In this case, the available bandwidth is sufficient to transmit all locally detected variables, and the semantic selection of these variables has no impact on the average semantic value $\overline{SV}$.

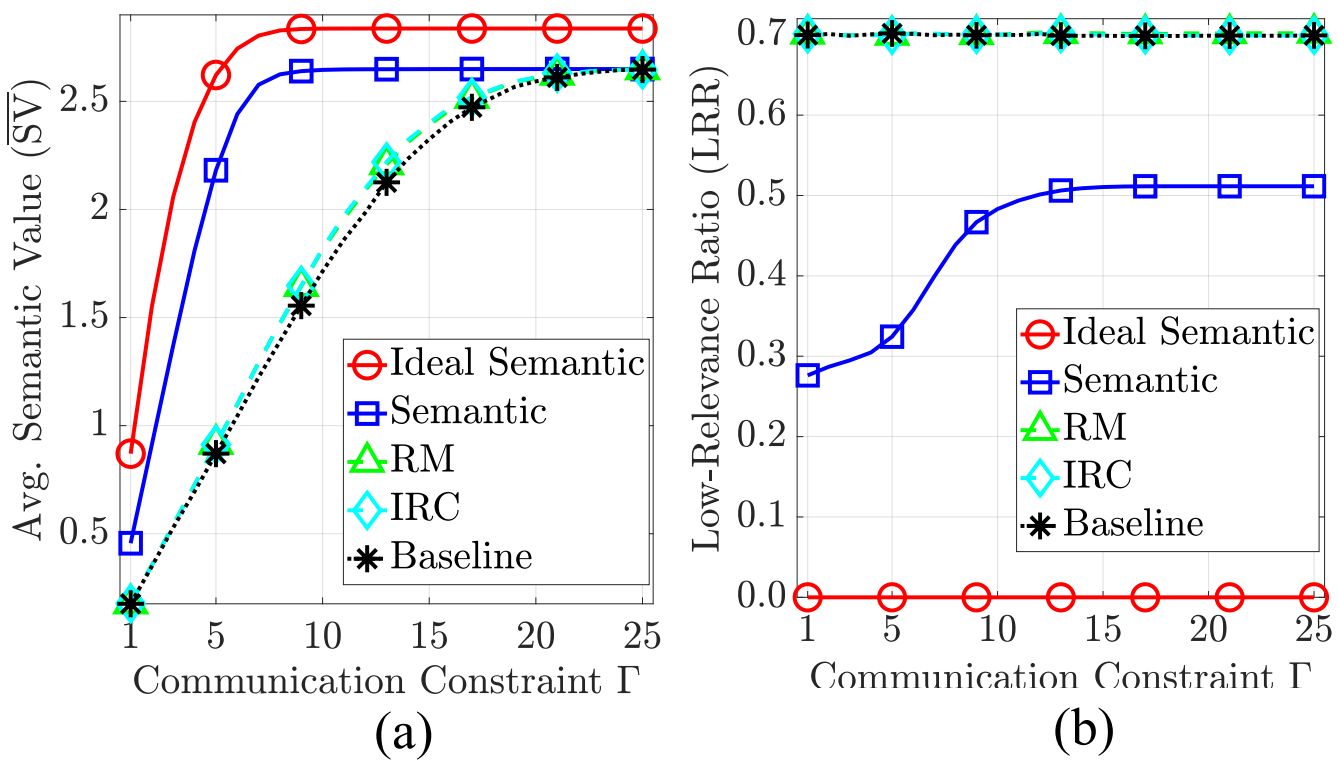

**Fig. 6.** $\overline{SV}$ (a) and LRR (b). Unicast scenario.

Fig. 6(a) shows that existing V2X schemes (*Baseline, IRC,* and *RM*) can achieve the same average semantic value $\overline{SV}$ as the *Semantic* scheme when the communication constraint Γ is relaxed. However, this is accomplished with a much less efficient use of the communication channel. Fig. 6(b) reports the Low-Relevance Ratio (LRR) metric, which measures the fraction of irrelevant or low-relevance exogenous variables included in each transmitted message, as a function of the communication constraint Γ. In this study, an exogeneous variable is considered irrelevant if its context-dependent semantic value is below a threshold $w_{min}$, set at 0.05. Fig. 6(b) shows that the *Ideal Semantic* and *Semantic* schemes consistently outperform the *RM, IRC,* and *Baseline* schemes across all values of Γ. The semantic schemes consistently transmit a much smaller fraction of irrelevant variables, resulting in a more efficient use of communication bandwidth. This holds true even for Γ values equal to or greater than 20, which is the threshold at which the *Baseline, IRC,* and *RM* schemes reach the same average semantic value $\overline{\text{SV}}$ as the *Semantic* scheme in Fig. 6(a). However, they do so at a significantly higher communication cost, as *Baseline, IRC,* and *RM* transmit a considerably larger fraction of irrelevant variables. Fig. 6(b) further highlights the importance of improving the estimate of the contextual relevance function of the intended receiver to unlock the full potential of semantic and task-oriented communications. The figure shows that the *Ideal Semantic* scheme consistently achieves LRR values of zero across all Γ values by perfectly identifying and filtering irrelevant information. Although the *Semantic* scheme significantly reduces LRR compared to the *Baseline, IRC,* and *RM* schemes, it cannot bring LRR to zero due to an imperfect estimation of the contextual relevance functions. These inaccuracies prevent the *Semantic* scheme from identifying all low-relevance exogenous variables, leading it to transmit more irrelevant information than the *Ideal Semantic* scheme despite achieving comparable values of HRR (Fig. 5) and $\overline{\text{SV}}$ (Fig. 6(a)). These results demonstrate that the efficiency of semantic and task-oriented V2X communications strongly depends on accurately estimating the relevance of the information for the intended receiver.

By prioritizing the transmission of the most relevant information, the semantic schemes increase the average semantic value of transmitted messages (Fig. 6(a)) and reduce the number of transmitted variables. This is illustrated in Fig. 7(a), which presents the average number of exogenous variables $x_k$ per transmitted message ($|\overline{s^{ex}_{T,tx}}|$, a proxy of the message size), normalized with respect to the communication constraint Γ. The ratio $|\overline{s^{ex}_{T,tx}}|/\Gamma$ represents the communication resources' usage or the fraction of available communication resources utilized by each scheme to transmit each message. Fig. 7(a) demonstrates that the *Ideal Semantic* and *Semantic* schemes significantly reduce the number of transmitted variables compared to existing schemes (*Baseline, IRC,* and *RM*) by focusing on the transmission of the most relevant variables. This approach reduces bandwidth consumption and enhances communication efficiency. In contrast, the *Baseline* and *IRC* schemes consume – and waste – the largest amount of resources. The *Baseline* scheme attempts to transmit all locally detected exogenous variables and the *IRC* scheme removes only the subset of redundant variables needed to meet the constraint Γ. With respect to the *Baseline* and *IRC* schemes, the *RM* scheme reduces bandwidth consumption by filtering out all redundant variables. However, non-redundant variables can still be irrelevant, preventing *RM* from achieving the performance levels of semantic schemes. Consequently, *RM* transmits 3.9 and 1.6 times more exogenous variables than the *Ideal Semantic* and *Semantic* schemes, respectively, when $\Gamma > 20$ (Fig. 7(a)). The *Semantic* scheme cannot reach the values of $|\overline{s^{ex}_{T,tx}}|/\Gamma$ obtained with *Ideal Semantic* due its imperfect estimation of the contextual relevance function of the intended receiver, which leads to the transmission of irrelevant variables (Fig. 6(b)).

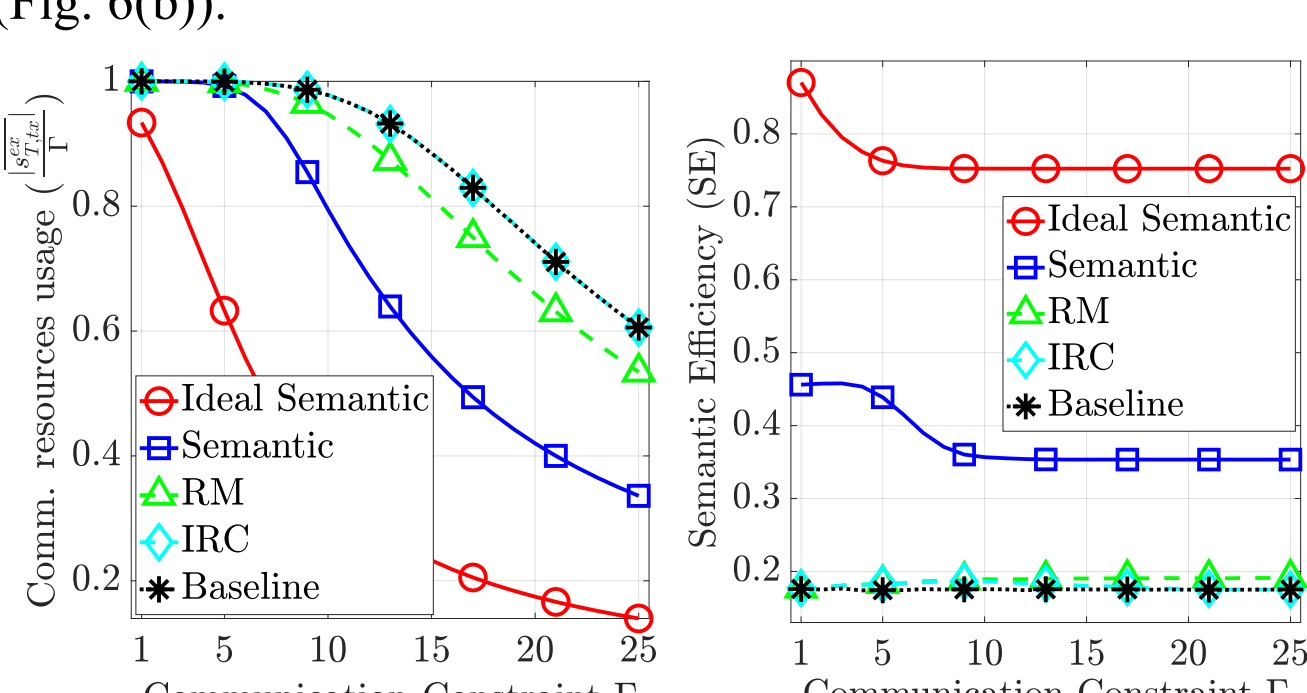

**Fig. 7.** $|\overline{s^{ex}_{T,tx}}|/\Gamma$ (a) and SE (b). Unicast scenario.

Fig. 5 through Fig. 7(a) numerically demonstrate the unique benefits of semantic and task-oriented V2X communications. Semantic and task-oriented V2X communications enable the transmission of a larger amount of relevant information to the intended receiver (Fig. 5), increase the semantic value of transmitted messages (Fig. 6(a)), and reduce the number of transmitted exogenous variables (Fig. 6(b) and Fig. 7(a)). In essence, these results demonstrate that semantic and task-oriented V2X communications can accomplish the communication goal more efficiently. Fig. 7(b) compares the efficiency of the five schemes reporting the Semantic Efficiency (SE) as a function of Γ. SE measures the average semantic value that is conveyed by each transmitted exogenous variable. This metric provides an indication of the communication efficiency as it captures the amount of resources that need to be consumed by each vehicle to convey the desired meaning that is relevant to the intended receiver. A lower SE means that vehicles must transmit a larger number of exogenous variables to accomplish the communication goal. Fig. 7(b) shows that, by focusing on the transmission of the most relevant information, semantic V2X schemes can augment the semantic efficiency with respect to relevance-agnostic schemes (*RM, IRC,* and *Baseline*), and optimize the bandwidth consumption. In particular, *Semantic* achieves a two-fold improvement in terms of semantic efficiency with respect to the *RM, IRC,* and *Baseline* schemes for all considered values of Γ.

It is important to note that these improvements are sensitive to key driving scenario parameters such as $\Delta_L$ and $p$, which define the fraction of relevant exogenous variables in each vehicle's contextual relevance function and the probability that the contextual relevance functions of neighbouring vehicles are uncorrelated, respectively. As $\Delta_L$ increases, the contextual relevance function of receiving vehicles features fewer relevant exogenous variables. Consequently, the *Semantic* scheme will reduce the number of locally detected variables transmitted by each vehicle, reducing the consumption of communication resources and further improving its semantic efficiency. Similarly, a smaller $p$ increases the likelihood that the same exogenous variable is relevant to multiple receivers. This allows a transmitting vehicle to provide its intended receivers with the required relevant information while sending less exogenous variables, hence reducing communication resources usage and further enhancing the *Semantic* scheme's semantic efficiency.

Furthermore, note that *RM*, *IRC*, and *Baseline* schemes achieve a similar SE across all Γ values, which indicates that avoiding (*RM*) or reducing (*IRC*) the transmission of redundant information is not sufficient to efficiently convey the desired meaning that is relevant to the intended receiver. The gap between the *Semantic* and *Ideal Semantic* schemes shows the importance of an accurate estimation of the contextual relevance function to improve the communication efficiency.

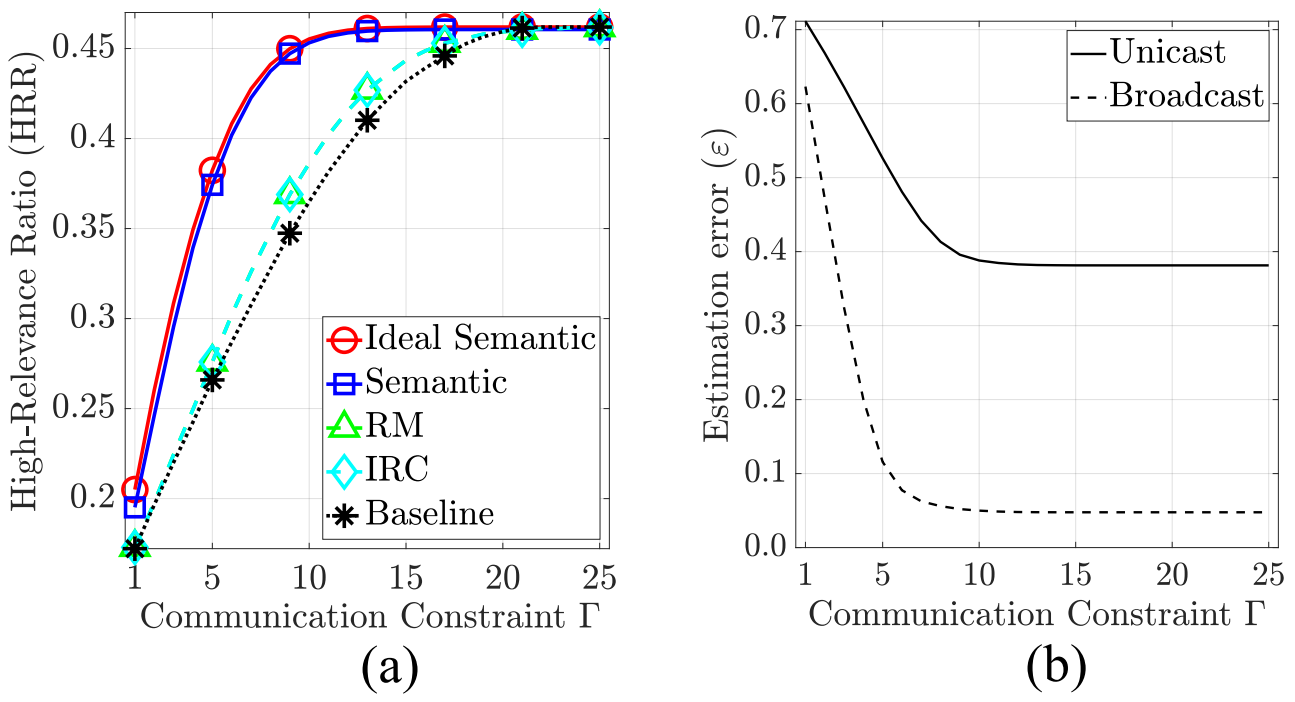

**Fig. 8.** (a) HRR. Broadcast scenario. (b) Relevance model's estimation error $\varepsilon$.

### *C. Broadcast*

We now analyze a broadcast communication scenario with $N$ = 4 communicating CAVs. Fig. 8(a) compares the fraction of high-relevance exogenous variables available at each vehicle (HRR) in the broadcast scenario. The figure shows that, like in the unicast scenario (Fig. 5), the *Ideal Semantic* and *Semantic* schemes transmit more relevant exogenous variables than the *RM, IRC,* and *Baseline* schemes, particularly at lower Γ values. At low Γ values, the *Semantic* scheme improvement with respect to relevant-agnostic schemes can be as large as 36%. These lower Γ values represent more constrained communication conditions, where a careful selection of transmitted information is crucial. The comparison between Fig. 5 and Fig. 8(a) reveals that an increase in the number of communicating CAVs particularly benefits the *Semantic* scheme, which achieves HRR values very close to the *Ideal Semantic* benchmark. As the number of communicating vehicles increases, the amount of contextual information exchanged over the V2X network also rises. This additional contextual information enables a more accurate understanding of the communication context, thereby reducing the relevance model's estimation error $\varepsilon$. This trend is evident in Fig. 8(b), which compares the estimation error $\varepsilon$ in unicast and broadcast scenarios as a function of Γ. The figure shows a notable reduction in $\varepsilon$ in the broadcast scenario, which improves the estimation of the intended receivers' contextual relevance function and explains why *Semantic* approaches the *Ideal Semantic*'s performance in Fig. 8(a).

The trends observed in Fig. 8(b) also explain the improvement in the average semantic value $\overline{\text{SV}}$ of the *Semantic* scheme in the broadcast scenario (Fig. 9(a)) compared to the unicast scenario (Fig. 6(a)). Fig. 9(a) compares the $\overline{\text{SV}}$ performance of the different V2X schemes in the broadcast scenario. By reducing the relevance model's estimation error, the *Semantic* scheme can better identify the most relevant exogenous variables and maximize the semantic value of transmitted messages in the broadcast scenario, achieving nearly the same $\overline{\text{SV}}$ performance as the *Ideal Semantic* benchmark. Fig. 9(a) also shows that the *Ideal Semantic* and *Semantic* schemes achieve higher $\overline{\text{SV}}$ values compared to the relevance-agnostic *RM, IRC,* and *Baseline* schemes in the broadcast scenario. Additionally, Fig. 9(a) reveals that increasing the number of communicating CAVs significantly deteriorates the $\overline{\text{SV}}$ performance of the *Baseline* and *IRC* schemes, especially at large Γ values, due to increased redundancy levels. A larger number of communicating vehicles increases the likelihood that an exogenous variable $x_k$ is redundantly detected by multiple vehicles due to overlapping local perception ranges. For the *Baseline* and *IRC* schemes, this leads to a higher probability that the transmitting vehicles include exogenous variables already available at the intended receivers, i.e., that are redundant. Recall that the *IRC* scheme does not remove all redundant variables from the transmitted messages, but only the subset needed to comply with the communication constraint Γ. The transmission of redundant exogenous variables reduces the semantic value per transmitted message, as we assume that the semantic value of a redundant exogenous variable $x_k$ is zero.

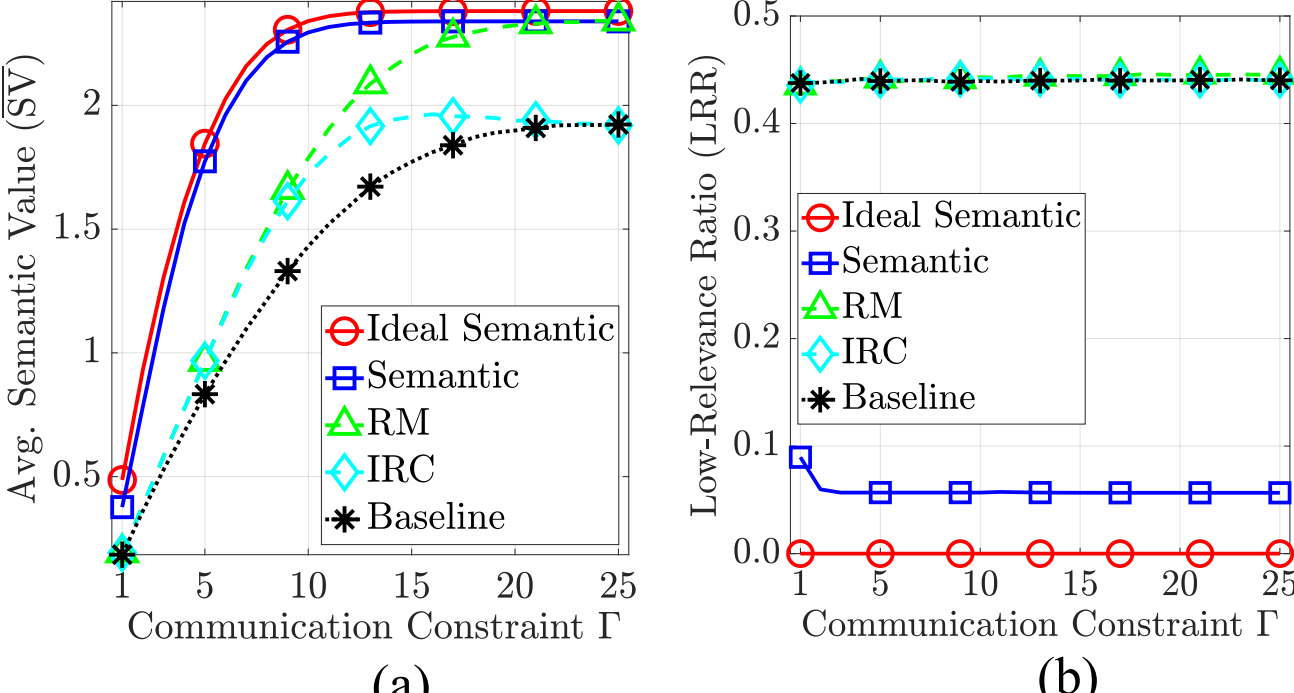

**Fig. 9.** $\overline{\text{SV}}$ (a) and LRR (b). Broadcast scenario.

The reduction of $\varepsilon$ observed in Fig. 8(b) also explains the reduction in LRR achieved by *Semantic* in the broadcast scenario (Fig. 9(b)) with respect to the unicast scenario (Fig. 6(b)). The *Semantic* scheme can achieve a more accurate estimate of the intended receivers' contextual relevance function in the broadcast scenario, and this helps it better identify the relevant (Fig. 9(a)) and irrelevant exogenous variables (Fig. 9(b)). We should note that *RM, IRC,* and *Baseline* also reduce the LRR in the broadcast scenario compared to unicast (0.7 to 0.43). This is because an exogenous variable is considered irrelevant in the broadcast scenario only when its semantic value is below $w_{min}$ for all intended receivers.

Fig. 10(a) analyzes the normalized average number of exogenous variables per transmitted message ($\overline{|s_{T,tx}^{ex}|}/\Gamma$) in the broadcast scenario. Comparing Fig. 7(a) (unicast) and Fig. 10(a) (broadcast) reveals that increasing the number of

communicating vehicles positively impacts the *Semantic* scheme's $|\overline{s_{T,tx}^{ex}}|/\Gamma$ performance. A larger number of transmitting vehicles improves the estimation of contextual relevance functions, and this helps the *Semantic* scheme to include only the most relevant exogenous variables in each message. For example, the $|\overline{s_{T,tx}^{ex}}|/\Gamma$ values achieved by the *Semantic* scheme at Γ = 10 and Γ = 15 decrease from 0.79 to 0.66 and from 0.55 to 0.45, respectively, when moving from the unicast to the broadcast scenario. Additionally, the gap in $|\overline{s_{T,tx}^{ex}}|/\Gamma$ between the *Semantic* and *Ideal Semantic* schemes is significantly reduced in Fig. 10(a), highlighting that semantic V2X can positively exploit an increasing number of communicating vehicles to optimize the message content. Note that the *Ideal Semantic* scheme exhibits higher $|\overline{s_{T,tx}^{ex}}|/\Gamma$ values in the broadcast scenario (Fig. 10(a)) than in the unicast one (Fig. 7(a)). For example, at Γ = 10 and Γ = 15, the *Ideal Semantic* scheme's $|\overline{s_{T,tx}^{ex}}|/\Gamma$ performance increases from 0.35 to 0.58 and from 0.23 to 0.39, respectively, when moving from unicast (Fig. 7(a)) to broadcast (Fig. 10(a)) communications. This occurs because broadcast communications involve a higher number of intended receivers, each with a different contextual relevance function. As a result, the *Ideal Semantic* scheme must include a larger set of exogenous variables in each transmitted message to ensure all the intended receivers obtain the most relevant information. The sensitivity of $|\overline{s_{T,tx}^{ex}}|/\Gamma$ to the number of intended receivers in broadcast communications (Fig. 10(a)) is not exclusive to the *Ideal Semantic* scheme but is an intrinsic characteristic of all semantic schemes. Fig. 10(a) also shows that *RM* reduces the $|\overline{s_{T,tx}^{ex}}|/\Gamma$ in the broadcast scenario, and augments its difference with the *IRC* and *Baseline* schemes. *RM* reduces the number of variables per transmitted message thanks to the increased redundancy levels experienced in the broadcast scenario. Nevertheless, the *RM* scheme cannot achieve the performance levels of the *Semantic* scheme. The performance gap between *RM* and *Semantic* scheme remains quite significant, with the *Semantic* scheme reducing the consumption of communication resources by up to 40%. The sensitivity of the *Ideal Semantic* scheme to the number of intended receivers increases the average number of exogenous variables per transmitted message in the broadcast scenario, reducing its communication efficiency. This effect is evident when comparing the semantic efficiency between the unicast (Fig. 7(b)) and broadcast (Fig. 10(b)) scenarios, with the *Ideal Semantic* average SE levels decreasing by approximately 50% in the broadcast scenario. In contrast, an increase in the number of communicating vehicles is especially beneficial for the *Semantic* scheme. With more contextual information available, the relevance model's estimation error decreases in the broadcast scenario, allowing the *Semantic* scheme to more accurately identify the most relevant exogenous variables and maximize the semantic value of the transmitted messages, while also reducing the number of exogenous variables per message. Improvements in $\overline{SV}$ and $|\overline{s_{T,tx}^{ex}}|/\Gamma$ performance offset the negative effects associated with the sensitivity of the *Semantic* scheme to the number of intended receivers, and its SE performance is not deteriorated in the broadcast scenario (Fig. 10(b)). The positive impact of an increasing number of communicating vehicles on the performance of *Semantic* is further demonstrated by the significant reduction in the SE gap between the *Semantic* and *Ideal Semantic* schemes when moving from the unicast (Fig. 7(b)) to the broadcast scenario (Fig. 10(b)).

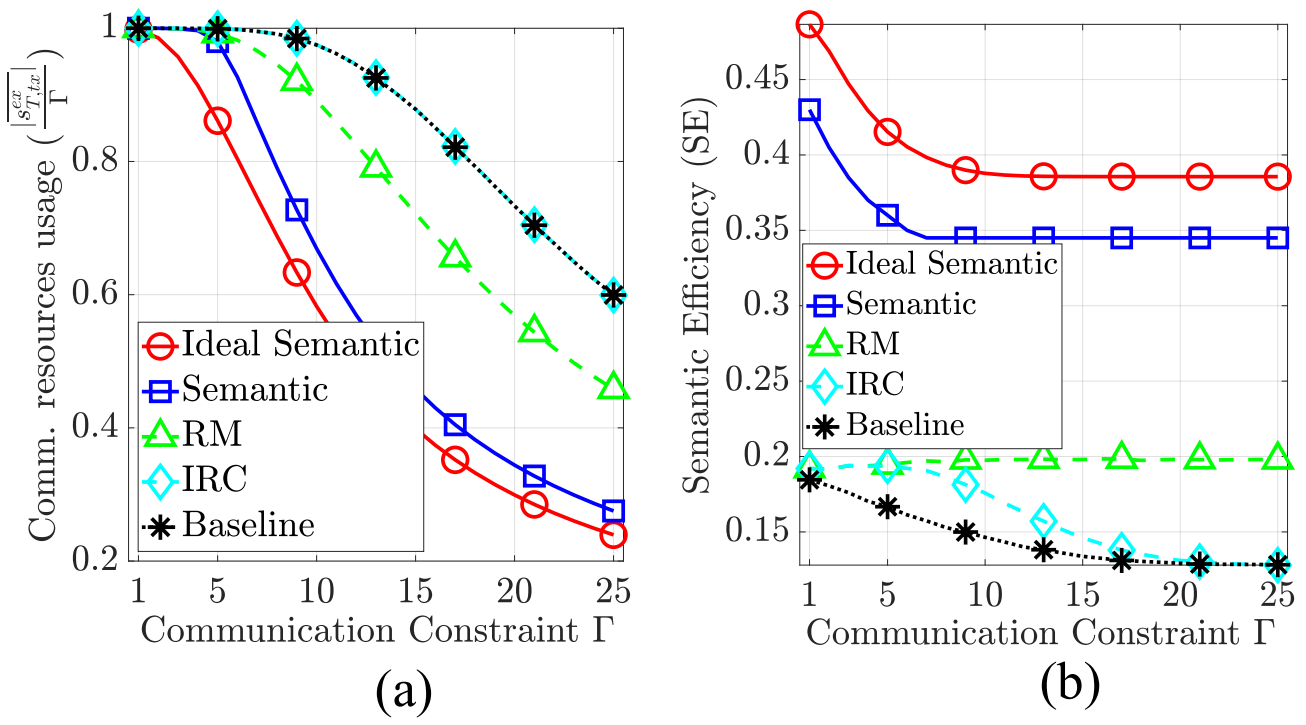

**Fig. 10.** $|\overline{s_{T,tx}^{ex}}|/\Gamma$ (a) and SE (b). Broadcast scenario.

The sensitivity of the semantic schemes to the number of intended receivers does not diminish their efficiency advantage over the relevance-agnostic *RM, IRC*, and *Baseline* schemes when transitioning from the unicast to the broadcast scenario. As shown in Fig. 10(b), the *Ideal Semantic* and *Semantic* schemes continue to outperform the *RM, IRC*, and *Baseline* schemes also in the broadcast scenario. The semantic schemes allow vehicles to communicate more efficiently by significantly reducing the average number of exogenous variables needed to maximize the semantic value of the transmitted messages and accomplish the communication goal. The *Ideal Semantic* and *Semantic* schemes achieve a 2x and 1.8x improvement in SE performance (hence, in communication efficiency), respectively, compared to the *RM* scheme in the broadcast scenario (Fig. 10(b)). Their gain over the *Baseline* and *IRC* schemes increases by up to 2.8x and 2.4x (Fig. 10(b)). It is worth noting that the increase in the number of communicating vehicles also benefits the *RM* scheme over the *Baseline* and *IRC* schemes in the broadcast scenario, as *RM* leverages the increased redundancy and reduces the average number of variables per message, ultimately increasing its SE (Fig. 7(b) vs Fig. 10(b)). In contrast, the *Baseline* scheme does not curate the content of the transmitted messages and the *IRC* scheme does not fully exploit the increased redundancy. As a result, their SE performance deteriorates as the number of communicating vehicles and redundancy levels increase.

### *D. Discussion*

By focusing on relevance, semantic and task-oriented V2X communications can prevent the transmission of unnecessary (or irrelevant) information and improve the communication efficiency. This is key to address the scalability challenges that V2X systems will face with the large-scale deployment of CAVs. However, estimating the context-dependent relevance of the transmitted information is not exempt of challenges due to the dynamism and complexity of the driving environment. Relevance estimation requires AI tools, computing platforms and algorithms that can interpret the available contextual information and robustly infer the intertwining between the context of a receiver and the relevance of a piece of information across different driving environments (e.g., urban, highway, etc.). These algorithms can be computationally intensive and, therefore, might require powerful hardware and software resources to guarantee real-time relevance estimation. However, CAVs are already semantic-based mobile platforms natively equipped with powerful hardware (CPUs and GPUs) and sophisticated AI algorithms. These capabilities are currently used by CAVs to semantically process raw data from onboard sensors and extract relevant information (e.g., lane margins, positions of surrounding objects) needed to plan and execute their driving maneuvers in real time. These hardware, software, and semantic processing capabilities can be leveraged for relevance estimation. They can be leveraged by vehicles to process and interpret contextual information, accurately estimate the intended receivers' context, and correctly identify the most relevant information in real-time, paving the way for the practical deployment of semantic and task-oriented V2X communications.

We should also note that the V2X domain represents a unique ecosystem for the realization and deployment of semantic and task-oriented V2X communications, not only due to its native semantic processing capabilities, but also to the inherent availability of rich contextual information. By design, the V2X domain functions as a native ISAC-type ecosystem, where CAVs cooperatively and continuously sense the surrounding environment and exchange locally generated information about the driving environment and their own mobility. For example, vehicles continuously share data about their position and speed, detected objects (e.g., vehicles, vulnerable road users), and even their driving intentions (e.g., planned trajectories). This data can be further aggregated, refined, and complemented by the roadside V2X infrastructure, which can provide additional information about the traffic conditions [2]. This rich contextual data is fundamental for an accurate understanding of the driving environment and context, and ultimately for estimating the relevance of information.

## VII. Conclusions

The large-scale deployment of CAVs will lead to a substantial increase in the volume of data exchanged over V2X networks, potentially challenging their communication efficiency and scalability. To address these challenges, this paper introduces a semantic and task-oriented communication paradigm where information is transmitted based on its relevance (or semantic value) for the intended receivers. We evaluate this paradigm in the V2X domain, an ecosystem with unique benefits for developing and deploying semantic and task-oriented communications. Our study demonstrates that semantic and task-oriented V2X communications can accomplish the V2X communication goal more efficiently than existing V2X solutions in both unicast and broadcast scenarios. By focusing on the transmission of relevant information, semantic and task-oriented V2X communications increase the semantic value of the messages delivered to the intended receivers while reducing the overall volume of transmitted data. In our evaluation, semantic and task-oriented V2X communications achieve nearly a two-fold improvement in communication efficiency by halving the communication resources needed by each transmitting vehicle to deliver the relevant information to the intended receivers. Our study also highlights the importance of accurately estimating the context-dependent relevance of information for intended receivers to fully harness the potential of semantic V2X communications, with notable differences observed between unicast and broadcast scenarios. Broadcast V2X communications increase the number of communicating vehicles but also enriches the amount of contextual information shared across the network. This additional information improves the accuracy of relevance estimation, enabling semantic and task-oriented V2X communications to effectively identify the most relevant information, maximize the semantic value of transmitted messages, and reduce bandwidth consumption. At the same time, broadcast V2X communications increase the number of intended receivers compared to unicast communications, potentially increasing the amount of information per message. However, the advantages of an increased availability of contextual information outweigh any drawbacks from serving multiple receivers, and semantic and task-oriented V2X communications are significantly more efficient than existing solutions in achieving V2X communication goals also in broadcast scenarios.

Our study has highlighted the potential of semantic and task-oriented V2X communications, while also identifying a major research challenge: accurately estimating the relevance of locally collected information for the intended receivers. It is worth highlighting that accurately estimating the relevance of locally collected information is not only fundamental for the deployment of semantic and task-oriented V2X communications, but it would also benefit autonomous driving systems more broadly. Current autonomous driving systems typically treat all objects in a scene equally, despite the fact that some objects are significantly more relevant to the driving task than others. Identifying these task-relevant objects would, for example, enable the development of more lightweight and accurate perception systems capable of prioritizing the detection and tracking of objects that have an impact on the autonomous vehicle's driving decisions. Given the broad benefits of relevance estimation, future research efforts should prioritize the development of the methodology and tools required to capture the intertwining between a vehicle's context and the relevance of a piece of information – namely, to construct real-world contextual relevance functions. The availability of these contextual relevance functions would then pave the way to the design and training of AI-based relevance models that CAVs can utilize to estimate, in real-time, the relevance of objects within the driving environment. Future work also includes analyzing the sensitivity of the proposed semantic and task-oriented V2X communication paradigm to different system settings, as well as evaluating its scalability in large-scale broadcast scenarios.